# Polaritonics in complex structures: Confinement, bandgap materials, and coherent control

**David W. Ward, Eric Statz, Jaime D. Beers, T. Feurer, John D. Joannopoulos\*, and Keith A. Nelson**
*Department of Chemistry and \*Center for Materials Science and Engineering¤, Massachusetts Institute of Technology, 77 Massachusetts Avenue, Cambridge, MA 02139*
**Ryan M. Roth and Richard M. Osgood**
*Microelectronics Sciences Laboratories, Columbia University, New York, NY 10027*
**Kevin J. Webb**
*School of Engineering and Computer Engineering,
Purdue University, West Lafayette, IN 47907*

**Abstract:** We report on the design, fabrication, and testing of ferroelectric patterned materials in the guided-wave and polaritonic regime. We demonstrate their functionality and exploit polariton confinement for amplification and coherent control using temporal pulse shaping.
©2003 Optical Society of America

Polaritonics is defined in an intermediate regime between electronics and photonics, roughly in the band above 100 GHz and below 10 THz. An ultrafast optical pulse focused into a ferroelectric crystal like $LiNbO_3$ (LN) or $LiTaO_3$ (LT) generates phonon-polaritons, admixtures of electromagnetic and TO phonon responses henceforth called polaritons, in this frequency range through impulsive stimulated Raman scattering (ISRS).[1] The polariton response is a traveling wave, so its propagation through the host crystal can be controlled using the techniques of guided wave, diffractive, and dispersive optics.[3,4] Additionally, temporal and spatio-temporal pulse shaping may be employed to tailor the polariton waveform through coherent control.[4] This THz generation and guidance scheme offers a fully integrated platform that supports generation and manipulation in a single patterned material. Here we extend the polaritonics platform to include single and coupled resonance structures. The resonant modes within the structures are illustrated and exploited for coherent control and amplification through temporal pulse shaping. We also characterize a planar waveguide from which polariton signal images can be recorded even though the sample thickness is only 10 μm.

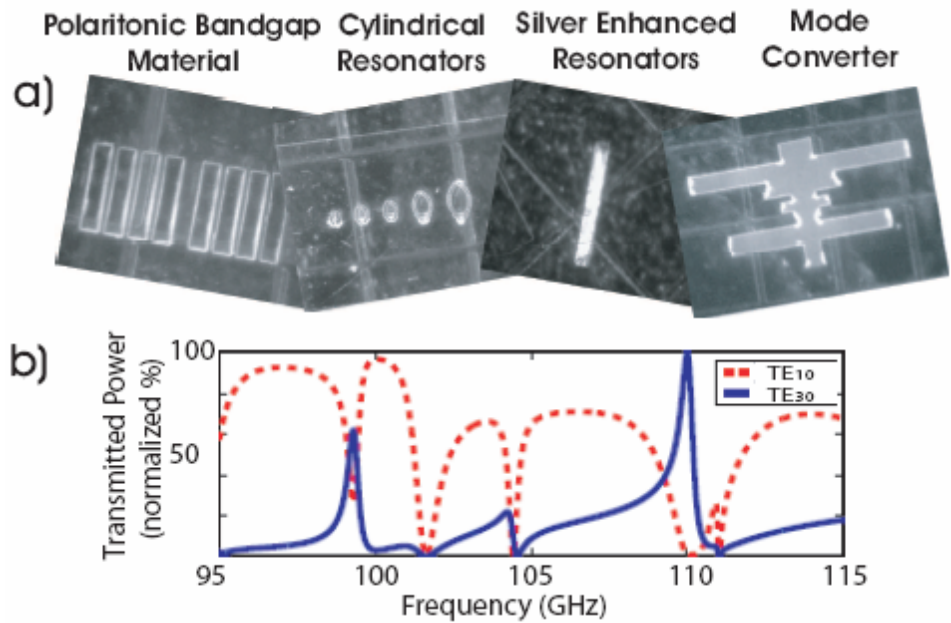

Fig. 1. (a)Machined structures including a 1-D polaritonic bandgap material, 2-D cylindrical resonators, silver coated resonators, and a mode converter structure. (b) The theoretical response of the pictured mode converter. With small changes in the polariton input frequency, the transmission properties of $TE_{10}$ and $TE_{30}$ modes change drastically.

## 1. Patterned Materials

Patterning of ferroelectric materials, which are resistant to conventional chemical processing, is achieved with ultrafast laser machining[5], which is conducted by focusing a series of ultrafast laser pulses (100-200 μJ, 50 fs) onto the crystal using a microscope objective (NA=1.4), leaving a material void in the irradiated region. In this manner 10-20 μm features that extend through the thickness (1-250 μm) of a LN or LT crystal are formed. A computer actuated Burleigh 3-axis translation stage controls the crystal position, allowing for the carving of user specified patterns into the crystal. Patterned materials fabricated in our group include resonators, waveguides, photonic bandgap materials, mode converters, as well as THz integrated platforms such as Mach-Zehnder interferometers, prisms, gratings, and waveguide couplers. Some of these structures are illustrated in figure 1a.

Presently, we are interested in structures that yield a high degree of specificity in the frequency dependence of distinct spatial modes. In the `polaritonic' bandgap regime, when the bandgap due to the spatial period overlaps spectrally with the intrinsic polariton bandgap, extreme localization of the electromagnetic energy and frequency dependent relocation may result.[6] Aperiodic structures also offer a means of sensitive frequency-dependent mode selection.[7] In this case, small changes in frequency result in nearly complete inter-conversion between longitudinal modes. Polaritons in the mode converter in fig 1a, for example, propagate in the $TE_{30}$ mode with 99% efficiency at 110 GHz and in the $TE_{10}$ mode at 100 GHz with 97% efficiency. The transmission and reflection spectra of this mode converter as calculated with finite element analysis are illustrated in fig 1b.

## 2. Ferroelectric Slab Waveguide

A 10 μm LN film was fabricated through ion implantation and peel-off.[8] To demonstrate its planar waveguide dispersion properties, we have generated narrowband polaritons through ISRS using crossed beam excitation[9] and monitored their temporal and spatial evolution using polariton imaging.[10] We observe that for large wavevector polaritons, in which the wavelength is still small in comparison to the crystal thickness, confinement effects are negligible and the resulting dispersive properties are the same as in bulk. At small wavevectors, in which the wavelength is larger than the crystal thickness, the dispersive properties deviate significantly from that in bulk. In this case, a significant portion of the energy density is exterior to the host crystal, and the polaritons experience an effective index that is an average of the bulk and cladding. The waveguide dispersion relation deviates increasingly from that of the bulk as the wavevector decreases, as illustrated in Figure 2.

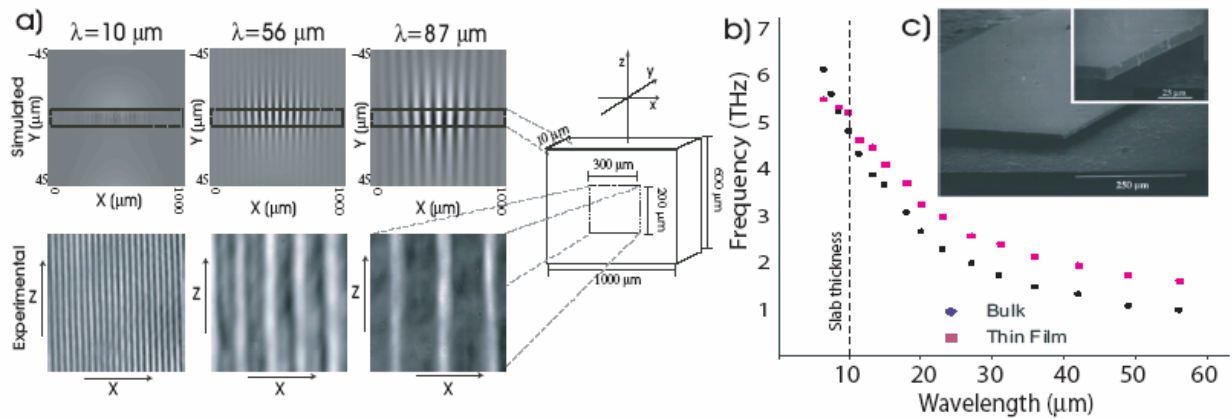

Fig. 2. Narrowband polaritons are generated using crossed excitation beams in 10 μm thick LN, (film shown in part c). Their spatial and temporal evolution are monitored simultaneously by polariton imaging (images from polaritons with 3 wavevectors are shown in part a), and the polariton dispersion curve results from extracting frequency and wavevector information from each scan. Deviations of the dispersion relation from that in bulk are apparent when the polariton wavelength is longer than the slab thickness, as indicated in part b.

## 3. Coherent Control

Coherent control over polaritons ordinarily requires spatial as well as temporal pulse shaping in order to enable manipulation following propagation away from the excitation region. Confinement within a resonator, however, causes polaritons to return repeatedly to the excitation region where they may be manipulated by successive excitation pulses that arrive there with specified timing. Here we demonstrate coherent amplification of selected modes in a coupled-resonator structure consisting of a ferroelectric cavity skirted by two resonant THz air cavities. The three coupled resonators are analogous to three coupled oscillators, and the perturbational resonance frequencies $\nu_{res}^{sym}$ are given in terms of the uncoupled resonant frequencies, $\nu_{LN}$ and $\nu_{air}$, and a coupling constant, $\kappa$, which is proportional to the index contrast and resonator widths: $\nu_{res}^{sym} = \frac{1}{2}(\nu_{LN} + \nu_{air}) \pm \left(\frac{1}{4}(\nu_{LN} - \nu_{air})^2 + 2\kappa^2\right)^{1/2}$. A third resonance frequency corresponding to an anti-symmetric mode whose energy is largely in the two air cavities is possible, but its excitation through ISRS and observation through LN index changes would be difficult.

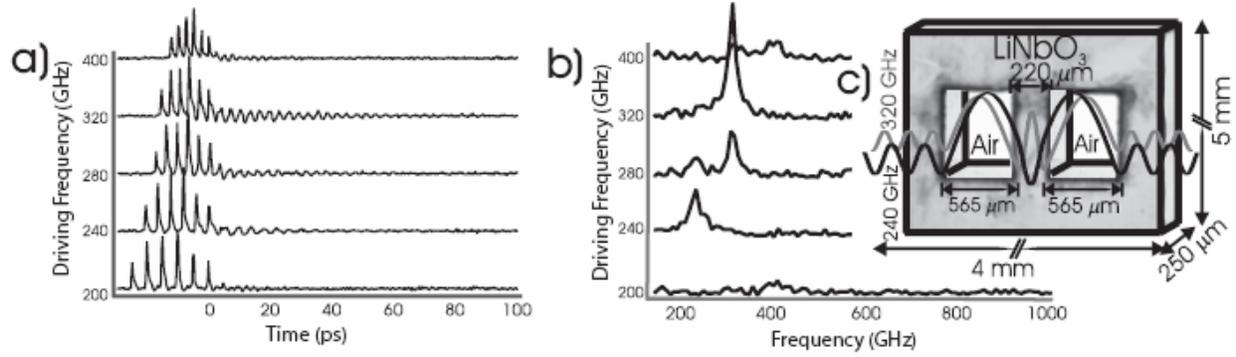

Fig. 3. Illustration of selective driving of the coupled-resonator resonant frequencies in the (a) time and (b) frequency domain. The prominent initial peaks in (a) are associated with both electronic (tall and sharp) and polariton (smaller and somewhat broader) responses. Polariton signals are strongest and longest in persistence when the cavity modes are driven on resonance. The resonant symmetric eigenmodes and relevant dimensions of the coupled resonator are indicated (c).

A 1 μJ excitation pulse is split into a train of six pulses with a nearly Gaussian profile using a recirculating reflective pulseshaper, which is capable of generating optical pulse-trains that are tunable in the 5 GHz - 5 THz range through the motion of a single delay stage. Figure 3 shows that either symmetric mode may be selectively driven and amplified by an appropriately tuned pulse sequence. Amplification by a factor of ten in energy has been observed.

## 5. Conclusion
In conclusion, we have extended the polaritonics platform to include single and coupled resonance structures and planar waveguides. The characteristic polariton responses in these structures have been characterized, and resonance has been exploited for coherent control by properly timed pulse sequences.

## 6. References

1. T.P. Dougherty, G.P. Wiederrecht, and K.A. Nelson, "Impulsive stimulated Raman scattering experiments in the polariton regime," J. Opt. Soc. Am. B **9**, 2179-2189 (1992).
2. N.S. Stoyanov, D.W. Ward, T. Feurer, and K.A. Nelson, "Terahertz polariton propagation in patterned materials," Nat. Mat. **1**, 95-98 (2002).
3. N.S. Stoyanov, T. Feurer, D.W. Ward and K.A. Nelson,"Integrated diffractive THz elements," Appl. Phys. Lett. **82**, 674-676 (2003).
4. T. Feurer, J.D. Vaughan and K.A. Nelson, "Spatiotemporal coherent control of lattice vibrational waves," Science, **299**, 374-377 (2003).
5. G.B. Schaffer, A. Brodeur, J.E. Garcia, and E. Mazur, "Micromachining bulk glass by use of femtosecond laser pulses with nanojoule energy," Opt. Lett. **26**, 93-95 (2001).
6. K.C. Huang, P. Bienstman, J.D. Joannopoulos, K.A. Nelson, and S. Fan, "Field Expulsion and Reconfiguration in Polaritonic photonic Crystals," Phys. Rev. Lett. **90**,196402 (2003).
7. M.C. Yang, J.H. Li, and K.J. Webb, "Functional field transformation with irregular waveguide structures," App. Phys. Lett. **83**, 2736-2738 (2003).
8. M. Levy, R.M. Osgood, Jr., R. Liu, E. Cross, G.S. Cargill III, A. Kumar and H. Bakhru, "Fabrication of Single-Crystal Lithium Niobate Films by Crystal Ion Slicing," Appl. Phys. Lett. **73**, 2293-2295 (1998).
9. T.F. Crimmins, N.S. Stoyanov, and K.A. Nelson, "Heterodyned impulsive stimulated Raman scattering of phonon-polaritons in LiTaO3 and LiNbO3," J. Chem. Phys. **118**,2882-2896 (2002).
10. R.M. Koehl, S. Adachi, and K.A. Nelson, "Direct visualization of collective wavepacket dynamics," J. Phys. Chem. A. **103**, 10260-10267 (1999).